\documentclass[fleqn,usenatbib]{mnras}
\usepackage{newtxtext,newtxmath}
\usepackage[T1]{fontenc}
\usepackage{ae,aecompl}
\usepackage{xcolor}

\usepackage{graphicx}	% Including figure files
\usepackage{amsmath}	% Advanced maths commands
\usepackage{amssymb}	% Extra maths symbols
\title[Faraday conversion in FRBs]{Faraday conversion and magneto-ionic variations in Fast Radio Bursts}

\author[Vedantham \& Ravi]{
H.~K.~Vedantham,$^{1}$\thanks{E-mail: vedantham@astron.nl}
V.~Ravi$^{2}$
\\
$^{1}$ASTRON, Netherlands Institute for Radio Astronomy, Oude Hoogeveensedijk 4, 7991PD, Dwingeloo, The Netherlands\\
$^{2}$Harvard-Smithsonian Center for Astrophysics, 60 Garden Street, Cambridge, MA 02138, USA
}

\date{Accepted XXX. Received YYY; in original form ZZZ}

\pubyear{2018}

\begin{document}
\label{firstpage}
\pagerange{\pageref{firstpage}--\pageref{lastpage}}
\maketitle
\begin{abstract}
The extreme, time-variable Faraday rotation observed in the repeating fast radio burst (FRB) 121102 and its associated persistent synchrotron source demonstrates that some FRBs originate in dense, dynamic and possibly relativistic magneto-ionic environments.  Here we show that besides rotation of the linear-polarisation vector (Faraday rotation), such media can generally convert linear to circular polarisation (Faraday conversion). We use non-detection of Faraday conversion, and the temporal variation in Faraday rotation and dispersion in bursts from FRB\,121102 to constrain models where the progenitor inflates a relativistic nebula (persistent source) confined by a cold dense medium (e.g. supernova ejecta). We find that the persistent synchrotron source, if composed of an electron-proton plasma,  must be an admixture of relativistic and non-relativistic (Lorentz factor $\gamma<5$) electrons. Furthermore we independently constrain the magnetic field in the cold confining medium, which provides the Faraday rotation, to be between 10 and 30\,mG. This value is close to the equipartition magnetic field of the confined persistent source implying a self-consistent and over-constrained model that can explain the observations.
\end{abstract}
\begin{keywords}
radio continuum: transients -- polarization -- radiative transfer
\end{keywords}
\section{Introduction}
By virtue of their large Faraday rotation measures, at least two Fast Radio Burst (FRB) sources \citep[FRB\,121102, FRB\,110523;][]{spitler2014,masui2015} are observed to reside in dense magneto-ionic environments. 
In addition to Faraday rotation, in the presence of mildly relativistic plasma ($\gamma\gtrsim 3$ typically), propagation through a magneto-ionic medium leads to Faraday conversion wherein linearly polarised light is converted to circularly polarised light and vice-versa \citep{sazonov1969, pachol1970, huang2011}. 
Faraday conversion is insignificant in the presence of typical interstellar magnetic fields %(where $\nu_{\rm B}\sim 10\,$Hz)
, but is thought to result in significant circular-polarisation fractions observed in Active Galactic Nuclei jets \citep{hofman2001} and in Sgr\,A$^\ast$ \citep{bower2002}, where the radiation propagates through a relativistic media with much larger magnetic fields than the Milky Way interstellar medium.

In this paper, by taking FRB\,121102 as a test case, we argue that Faraday conversion is an observable effect in some FRBs and leads to upper limits on the circum-burst magnetic field and density of low Lorentz factor electrons ($3\lesssim \gamma\lesssim 100$) that are otherwise inaccessible. We show that additional constraints from temporal variations in the dispersion and Faraday rotation of FRB\,121102 critically constrain proposed models for the environment of FRB\,121102. 

In \S\ref{sec:obs_cons}, we summarise existing observations and model-independent constraints on the magneto-ionic environment of FRB\,121102  and its associated persistent radio source. In \S\ref{sec:fc_limits}, we  describe the additional constraints implied by the non-detection of circular polarisation in FRB\,121102, given predictions from Faraday conversion. We discuss the implications of these results for models where FRB\,121102 is located within the persistent radio source in \S\ref{sec:discussion}, and conclude in \S\ref{sec:conclusions}.

\section{General constraints on the environment of FRB 121102}
\label{sec:obs_cons}

\subsection{Observations}
We first summarise the known properties of the magneto-ionic environment surrounding the repeating FRB\,121102 \citep{spitler2016}. It has been localised to an HII region in a galaxy at a redshift  $z=0.19$ \citep[luminosity and angular-diameter distances of $d_{\rm L} \approx 970\,$Mpc and $d_{\rm A}\approx 680\,$Mpc respectively;][]{chatterjee2017}.
Additionally, the FRB\,121102 bursts are (i) co-located to within $40\,$pc (95\% confidence) with  a persistent flat-spectrum ($1-10$\,GHz) radio source  with flux-density $S_\nu \approx 200\,\mu$Jy at $3\,$GHz \citep{marcote2017},  and (ii) show very high levels of Faraday rotation, quantified by the rotation measure, ${\rm RM}=1.46 \times 10^5\,{\rm rad}\,{\rm m}^{-2}$ in the source frame. The RM reduced by $\sim10$\% in seven months \citep{michilli2018}, whereas its dispersion measure increased by just $1-3\,{\rm pc}\,{\rm cm}^{-3}$ in four years \citep{hessels2018}.

\subsection{Dispersion and Faraday rotation}
\label{subsec:dm_rm}
The dispersion measure in the entire host galaxy is constrained to be be less than $250\,{\rm pc}\,{\rm cm}^{-3}$ \citep{tendulkar2017}. If an amount, ${\rm DM}_{\rm RM}$, of that is in the Faraday rotating nebula, \citet{hessels2018} obtain the following bound by requiring the magneto-ionic medium to be transparent to free-free absorption at the lowest frequency at which bursts have been observed (1\,GHz):
\begin{equation}
T_4^{2.3}{\rm DM}_{\rm RM}>150\,\left(\frac{\beta}{\eta_B^2}\right)
\end{equation}
where $T_4$ is the gas temperature in units of $10^4\,$K, $\beta$ is the ratio of thermal to magnetic pressure, and $\eta_B\leq 1$ is a geometric factor equal to the mean value of $\cos\theta$ along the ray path in the Faraday rotating medium where $\theta$ is the angle the ray makes with the ambient magnetic field. For an ordered field, we have $\eta_B=\cos\theta$. Significantly smaller values must be expected for a highly tangled field due to partial cancellation of positive and negative Faraday rotation. Further, for any given choice of ${\rm DM}_{\rm RM}$, the corresponding magnetic field that can generate the observed Faraday rotation is
\begin{equation}
\label{eqn:b_dm}
B = \frac{0.18}{{\rm DM}_{\rm RM}\eta_B}\,\,{\rm G}
\end{equation}
We now address the RM variations using two generic models for the magneto-ionic medium. We will address specific models in \S\ref{sec:discussion}.

{\em Expanding nebula}: If the observed RM decrease is due to the expansion of a nebula  of radius $R$ \citep[e.g.,][]{waxman2017,margalit18}, then ${\rm DM}_{\rm RM}\propto R^{-2}$ and $B\propto R^{-1.5}$, maintaining the same plasma $\beta$ and geometric factor $\eta_B$. We therefore have $\Delta{\rm RM}/{\rm RM} = \Delta B/B + \Delta {\rm DM}_{\rm RM}/{\rm DM}_{\rm RM}$, where (to first order) $\Delta B/B =-1.5\Delta R/R$, and $\Delta {\rm DM}_{\rm RM}/{\rm DM}_{\rm RM} = -2\Delta R/R$. The observed 10\% variation in RM over seven months then implies $\Delta R/R\approx 0.05\,$yr$^{-1}$ and the nebula is therefore expanding on a timescale of about $\tau\approx 20$\,years. The implied $-5.8$\% {\em decrease} in ${\rm DM}_{\rm RM}$ due to the expansion must be insignificant compared to the observed {\em increase} of $1-3\,{\rm pc}\,{\rm cm}^{-3}$ in the total DM over a longer timescale than the reported RM variations; the observed DM variations presumably occur in plasma unrelated to the Faraday-rotating plasma. Hence we can safely place the constraint ${\rm DM}_{\rm RM}< 17.5\,{\rm pc}\,{\rm cm}^{-3}$.
These constraints do not differ significantly for other expansion scenarios. For instance, adiabatic expansion of a tangled field ($B\propto R^{-2}$), gives $\tau\approx 23.33\,$yr and ${\rm DM}_{\rm RM}<20\,{\rm pc}\,{\rm cm}^{-3}$.

{\em Transverse motion}:
Alternatively, the RM variations could be due to transverse motion of the FRB source across a nebula over a characteristic timescale of $\tau \approx 5.83\,$yr. Let the line-of-sight extent of the Faraday-rotating nebula be equal to its transverse scale length. Then we have $\Delta{\rm RM}/{\rm RM} \approx \Delta {\rm DM}_{\rm RM}/{\rm DM}_{\rm RM}\approx 0.1$. The lack of significant DM variations accompanying the RM variations now gives a more stringent constraint of ${\rm DM}_{\rm RM}< 10\,{\rm pc}\,{\rm cm}^{-3}$.

Using the formalism from \S4.4 of \citet{hessels2018}, we can translate the above limits on ${\rm DM}_{\rm RM}$ to the underlying plasma parameters (${\rm DM}_{10} = {\rm DM}_{\rm RM}/(10\,{\rm pc}\,{\rm cm}^{-3})$ hereafter):
\begin{eqnarray}
\label{eqn:dm_rm_limits}
{\rm DM}_{10} &<&(1.75,\, 1.0) \nonumber \\
T_4&>&2.72\,\left(\frac{\beta}{\eta^2_B}\right)^{1/2.3}\,{\rm DM}_{10}^{1/2.3} \nonumber\\
B&=&18/({\rm DM}_{10}\eta_B) \,\,{\rm mG} \nonumber \\
n_e&=&9.3\times 10^6\,\left(\frac{\eta_B^2T_4}{\beta}\right)^{-1}\,{\rm DM}_{10}^{-2} \,\,{\rm cm}^{-3} \nonumber \\
l&=&1.1\times 10^{-6}\,\left(\frac{\eta_B^2 T_4}{\beta}\right)\,{\rm DM}_{10}^3 \,\,{\rm pc}\nonumber \\
\tau & = & (20,\,5.83)\,\,{\rm yr}\nonumber \\
v &=& (0.054,\,0.185)\,\left(\frac{\eta_B^2 T_4}{\beta}\right)\,{\rm DM}_{10}^3\,\,{\rm km}\,{\rm s}^{-1}
\end{eqnarray}
Here the two limits within parentheses are for the expanding-nebula and transverse-motion scenarios respectively, and $l$ and $n_e$ are the thickness and electron density of the Faraday-rotating medium respectively. The velocity, $v$, in the two cases must be interpreted as the radial expansion speed of the nebula and transverse speed of the source with respect to the Faraday rotating medium, respectively. We will return to these constrains with regards to specific models in \S 4.

\subsection{The persistent radio source}
\label{subsec:persistent_source}
The properties of the persistent radio source associated with FRB\,121102 may be constrained independently of the Faraday-rotating medium. We assume equipartition between the relativistic gas and magnetic field as is common in synchrotron sources\footnote{In the model of \citet{waxman2017}, which we discuss in \S \ref{subsec:waxman_model}, equipartition is required by dynamical and source-size (from scintillation) constraints.} \citep{readhead1994}. The source becomes self-absorbed at $1.5\,$GHz for radius $R_{\rm per}< 0.05\,$pc; this is thus the lower bound on the source size. European VLBI Network observations of the source at 5\,GHz set an upper bound on the source radius of $R_{\rm per}\lesssim0.35$\,pc \citep{marcote2017}.  This is consistent with the $\approx 30\%$  amplitude modulations observed in the source at 3\,GHz \citep{chatterjee2017} being caused by refractive interstellar scintillation in the Milky Way ISM \citep{walker1998}. 
%its size must be much smaller than the refractive scale, which in the direction of FRB\,121102 is about 1\,mas or 3\,pc \citep{walker1998}.  
For any radius within the allowed range ($0.05<R_{\rm per}/{\rm pc}<0.35$), we can determine the equipartition magnetic field, $B_{\rm eq}$, and the column of relativistic electrons, $N_{\rm rel}$, using the standard expressions for synchrotron emissivity and absorption coefficients \citep[][their eqns. 6.36 \& 6.53]{rybicki_lightman}. We assume a power-law energy distribution of radiating electrons with somewhat shallow index of $b=-1.5$ that can account for the relatively flat spectrum of the source \citep{chatterjee2017}. The peak Lorentz factor of the distribution, $\gamma_{\rm max}$ is chosen to correspond to the observed spectral break frequency of $\nu_{\rm max}=10\,$GHz. If the lower Lorentz factor cut-off corresponds to emission at $\nu_{\rm min}=1\,$GHz\footnote{This assumption will be relaxed in \S\ref{sec:faraday_conversion}} then  the equipartition magnetic field and electron column thus determined for minimum and maximum source sizes are: $B_{\rm eq}\approx 140\,$mG, $\gamma_{
\rm min}\approx 50$, $\gamma_{\rm max}\approx 160$, $N_{\rm rel} \approx 0.95\,{\rm pc}\,{\rm cm}^{-3}$ for $R_{\rm per}=0.05\,$pc, and $B_{\rm eq}\approx 27\,$mG, $\gamma_{
\rm min}\approx 120$, $\gamma_{\rm max}\approx 370$, $N_{\rm rel} \approx 0.1\,{\rm pc}\,{\rm cm}^{-3}$ for $R_{\rm per}=0.35\,$pc. The reader can scale the equipartition field to other source sizes using $B_{\rm eq}(R)\propto R^{-6/7}$. The total energy contained in the relativistic electrons and the magnetic field (`equipartition energy'), is $\sim 10^{49.1}$ and $\sim 10^{50.2}$\,erg respectively.  If the relativistic electrons were injected in a one-off event, the synchrotron cooling rates at $\gamma_{\rm max}$ yield source ages of $14\,$yr for $R=0.05$\,pc and 60\,yr for $R=0.35\,$pc. The corresponding expansion velocities are $0.011\,c$ and $0.02\,c$ respectively.

\section{Faraday conversion}
\label{sec:faraday_conversion}
\begin{figure}
\centering
\includegraphics[width=\linewidth]{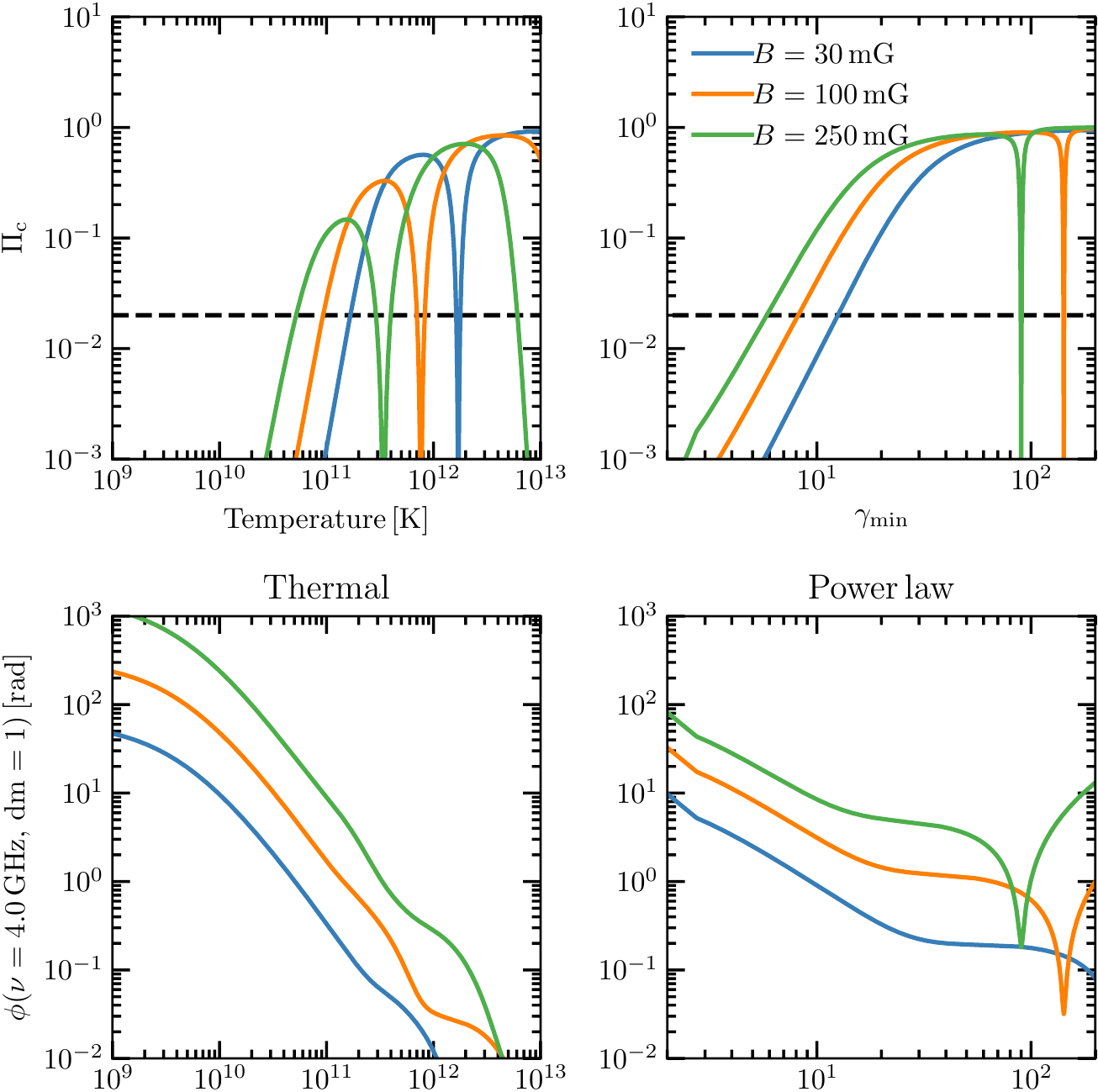}
\caption{The effect of elliptical birefringence at $\nu=4\,$GHz for thermal (left panels) and power law electron populations (right panels; energy index of $-1.5$), for three different magnetic field values (different line colours). Top panels show the peak circular fraction and bottom panels show the phase angle associated with generalised Faraday rotation. An electron column of $N_e=1\,{\rm pc\,cm}^{-3}$ (which is 1 DM units) has been assumed in all plots. Dashed black line in the top panels is placed at 2\% which corresponds to the observed linear fraction of $\gtrsim 98$\% in FRB\,121102 \citep{michilli2018}. The deep notches in the plots are due to zero-crossings. \label{fig:fc_rel}}
\end{figure}
\label{sec:fc_limits}

If the FRB progenitor resides within the persistent source, then relativistically corrections to the effects of birefringence must be considered to derive the levels of Faraday rotation and conversion (collectively called generalised Faraday rotation) in the persistent source. The effect is readily visualised on the Poincar\'{e} sphere as the rotation of the polarisation vector about an axis defined by the natural modes in the medium \citep[see ][for details]{kennett1998}. Two conditions must be satisfied to attain appreciable conversion of linear to circular polarisation due to propagation effects: (a) the natural modes in the medium must be sufficiently elliptical as characterised by their axial ratios, and (b) there must be a sufficient magneto-ionic column for this elliptical birefringence to have a measurable effect.
In Fig. \ref{fig:fc_rel}, we plot the level of generalised Faraday conversion for two commonly encountered electron distributions: a relativistic Maxwellian and a power law with an assumed index $b=-1.5$. We use the approximate expressions of \citet[][their eqns. 51, 58 \& 59]{huang2011} to do so. The upper panels show the peak circular fraction allowed by the ellipticity of the natural modes (condition (a) above) and the bottom panels show the generalised Faraday rotation angle (condition (b) above). The plots assume $\nu=4\,$GHz, $\theta=\pi/4$ (giving $\eta_B\approx 0.707$), and are normalised to a total electron column of $1\,{\rm pc}\,{\rm cm}^{-3}$. It is worth noting that in the power-law case, the mode ellipticity goes from the cold-plasma limit (circular modes) to its ultra-relativistic limit (linear modes) in a rather narrow range of $2\lesssim \gamma_{\rm min}\lesssim 20$--- a range that is practically inaccessible to photometric observations.

For `one-zone' models where the synchrotron emission and Faraday rotation come from the same nebula, condition (b) above is satisfied by definition and condition (a) must be reconciled with the non-detection of circular polarised emission. If the electron energies are power law distributed, then Fig. \ref{fig:fc_rel} (top-right panel) shows that for $B=30,\,100\,\&250\,$G, the circular fraction is in tension with observations for $\gamma_{\rm min}>3.6,\,2.3\,\&1.7$ respectively. The one-zone nebula must therefore be an admixture of synchrotron electrons ($50\lesssim \gamma \lesssim 370$) and `cold' plasma ($\gamma\lesssim 3$).

For `two-zone' models, bulk of the observed Faraday rotation occurs outside the synchrotron source in presumably cold plasma that does not yield significant conversion to circular polarisation. If the radio bursts originate from within the synchrotron source, then reconciling with observations requires one to ensure that both condition (a) and (b) are not simultaneously satisfied in the synchrotron source itself. Taking equipartition solutions for the persistent source from \S\ref{subsec:persistent_source} with $\nu_{\rm min}=1\,$GHz, we find that the model is in tension with the observations for $R<0.31\,$pc. If we allow $\gamma_{\rm min}$ to correspond to $\nu_{\rm min}=100\,$MHz, then the equipartition solutions are in tension with polarimetric data over the entire feasible parameter range of $0.05<R/{\rm pc}<0.35$. However, by extending the energy distribution to $\gamma_{\rm min}\lesssim 3$ (and admixture of `cold' and relativistic electrons), the modes can be constructed to be sufficiently circular so as to produce $<2$\% circular polarisation as in the `one-zone' case.

In summary, in all models where the radio bursts pass through the persistent source powered by a power-law electron energy distribution, the distribution must extend to $\gamma_{\rm min}\lesssim 3$, failing which (i) the Faraday screen cannot be co-located with the synchrotron emitting electrons, and (ii) the synchrotron source must have a radius in a narrow range of $0.31<R/{\rm pc}<0.35$. If the electrons are all injected into the nebula are highly relativistic, then the $\gamma_{\rm min}<3$ can be attained by radiative cooling over a timescale of $275$ and $7400\,$yr for $R=0.05$ and $R=0.35\,$pc respectively. If the electrons instead cool by adiabatic expansion from an injection Lorentz factor of $\gamma\gtrsim \gamma_{\rm max}$, then they must have been injected at when the nebula was $<(9.4\times 10^{-4},\,2.8\times 10^{-3})\,$pc if the present size of the nebula is $(0.05,0.35)\,$pc. These results can be directly applied to the one-zone magnetar model of \citet{margalit18}. Consider their benchmark model with $B=0.25\,$G, injection energy of $\gamma_{\rm inj} = 200$, energy distribution of $N_\gamma\propto\gamma^{-1.3}$ for $\gamma\leq \gamma_{\rm inj}$ and nebular age of $\tau = 12.4\,$yr. Lack of observed circular polarisation at 4\,GHz then requires $\gamma_{\rm min}<1.45$. If this is accomplished via adiabatic expansion then the electron injection must have started when the central source was $12.4\times 1.45/200= 0.09\,$yr. Faraday conversion constraints therefore require significant magnetic and baryonic flux to be ejected from the magnetar within a month of its birth. 

\section{Discussion}
\label{sec:discussion}

\subsection{Expanding relativistic nebula}
\label{subsec:waxman_model}
\begin{figure}
\centering
\includegraphics[width=\linewidth]{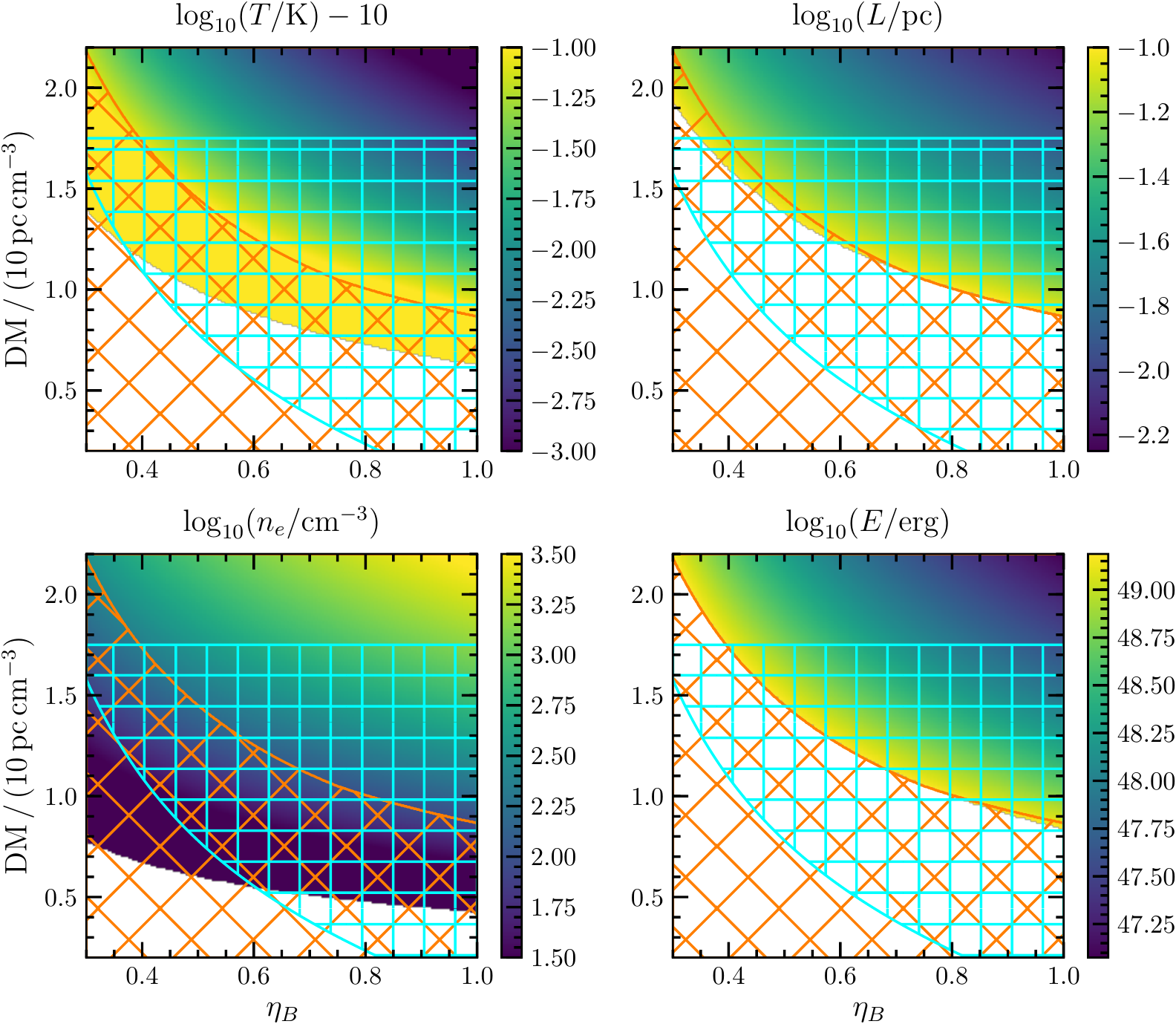}
\caption{Constraints on the expanding nebula model of \citet{waxman2017} determined from eqn.\ref{eqn:hot_nebula_params}, for plasma $\beta=1$ and persistent source radius of $R_{\rm per}=0.1$. The cross-hatched parameter space (orange) is excluded primarily due to energetic considerations (See \S\ref{sec:discussion}). The straight-hatched region (cyan) is the allowed parameter range due to constraints from Faraday conversion and magneto-ionic variations (see \S \ref{sec:fc_limits}) \label{fig:hot_nebula_constraints}}
\end{figure}

We consider the generic `two-zone' model of \citet{waxman2017}, where the source of FRB\,121102 is a compact object centrally located in a synchrotron nebula. The synchrotron nebula is confined against its tendency to relativistically expand by a much denser cold nebula. The Faraday rotation is provided by the shocked (and heated) part of the confining dense nebula.
Notwithstanding the FRB generating mechanism and the nature of the compact object, this model links the velocity of the shock driven by the expanding synchrotron nebula into the surrounding colder medium and the density of the latter medium via: $v_{\rm sh}\approx \sqrt{P_{\rm sh}/(n m_p)}$, where $m_p$ is the proton mass, and $P_{\rm sh} = B^2(1+\beta)/(8\pi)$ is the pressure in the shocked part of the nebula which is similar to that in the synchrotron source. In convenient units, we have
\begin{equation}
    \label{eqn:vshock}
    v_{\rm sh} \approx 9.1\left(\frac{T_4(1+\beta)}{\beta}\right)^{1/2}\,\,{\rm km\,s}^{-1}.
\end{equation}
We now equate $v_{\rm sh}$ with the expansion velocity in eqn. \ref{eqn:dm_rm_limits} to obtain the following family of models for the Faraday screen:
\begin{eqnarray}
\label{eqn:hot_nebula_params}
T_4 & = & 2.94\times 10^4\,\beta(1+\beta)\,\eta_B^{-4}\,{\rm DM}_{10}^{-6}\nonumber \\
l & = & 0.0323\,{\rm DM}_{10}^{-3}\,\eta_B^{-2}\,(1+\beta)\,\,{\rm pc} \nonumber \\
n_e & = & 316.3\,{\rm DM}_{10}^4\,\eta_B^2\,(1+\beta)^{-1}\nonumber \\
E_{\rm sh} & = & 10^{49.2}\,\left(R_{\rm per}/{\rm pc}\right)^2\,{\rm DM}_{10}^{-5}\,\eta_B^{-4}\,(1+\beta)^2\,\,{\rm ergs},
\end{eqnarray}
where $E_{\rm sh}$ is the combined thermal and magnetic energy in the shock-heated Faraday screen.
A feasible model from the above family must additionally satisfy the following constraints. (i) To avoid violating Faraday conversion constraints, we need $T_4<10^{5}\,$K (see Fig. \ref{fig:fc_rel} bottom-left panel). (ii) The Faraday-rotating plasma is presumably shock heated by the expanding relativistic gas, which requires the total energy in the latter to be larger than that in the former. (iii) As in \citet{waxman2017}, we assume that the thickness of the Faraday screen must be smaller than the radius of the persistent source. The constraints on the family of models are graphically shown in Fig. \ref{fig:hot_nebula_constraints} for a benchmark value of $R_{\rm per}=0.1\,$pc and $\beta=1$. We find feasible parameters ranges of ${\rm DM}_{10}\gtrsim 1.0$ and $\eta_B\gtrsim 0.4$ with a very weak dependence on $R_{\rm per}$. Taken together with the constraint of ${\rm DM}_{10}<1.75$ from \S\ref{subsec:dm_rm}, the allowed parameter ranges for the Faraday screen are tightly constrained:  ${\rm DM}_{10}\in[1.0,\,1.75]$, $\eta_B\gtrsim 0.4$, $B\eta_B\in[10,\,18]\,{\rm mG}$, and ${\rm log}_{10}(T/{\rm K})\in[7.5,\,9]$.  It is noteworthy and non-trivial that these self-consistent solutions should exist after a relationship between $v$, $T_4$ and $\beta$ due to eqn. \ref{eqn:vshock} was imposed on observational constraints from eqn. \ref{eqn:dm_rm_limits}.
\subsection{Dense filaments \`a-la Crab}
Following the suggestion by \citet{cordes2017} that FRB\,121102 is lensed by dense plasma structures similar to the cold filaments in the Crab nebula \citep{backer2000}, we consider a model where the variable RM and DM are obtained by transverse passage of dense filaments across the line of sight to the FRB source. \citet[][their \S 4.4 \& eqns. 6 to 9]{hessels2018} have summarised the resulting constraints in terms of the peak frequency at which lensing is apparent ($\nu_{\rm l}=8\,$GHz), the source$-$lens distance $D_{\rm sl}$ (units of pc), and the observer$-$lens distance $D_{\rm ol}\approx D_{\rm A}=622\,$Mpc. Requiring the filament to have the same transverse and line-of-sight extents, and to be transparent to free-free absorption at 1\,GHz, we obtain the following:
\begin{eqnarray}
\label{eqn:lensing}
T_4&>&5.5\,{\rm DM}_{10}^{1.15}\,D_{\rm sl}^{-0.385}\nonumber \\
n_e&>&2.7\times 10^6\,\left(\frac{{\rm DM}_{10}}{D_{\rm sl}}\right)^{1/2}\,\,{\rm cm}^{-3}\nonumber \\
\left(\frac{\eta_BT_4}{\beta}\right)&>&1.7\,D_{\rm sl}^{1/2}\,{\rm DM}_{10}^{-2.5} \nonumber \\
l&<&3.7\times 10^{-6}\,\left(D_{\rm sl}{\rm DM}_{10}\right)^{1/2}\,\,{\rm pc}
\end{eqnarray}

Anticipating large electron densities for filaments located within the persistent nebula ($d_{\rm sl}<1$), we impose an additional constraint: the Bremsstrahlung-cooling timescale should exceed $10\,$yr. Using \citet[][their eqn. 5.15b]{rybicki_lightman} the cooling-timescale is
\begin{equation}
\label{eqn:tau_ff}
    \tau_{ff} = 0.26\, T_4^{1/2}\,\left(\frac{n_e}{10^6\,{\rm cm}^{-3}}\right)^{-1}\,\,{\rm yr} > 10
\end{equation}
The simplest model we consider here is one where the Faraday rotation and lensing happens in the same filamentary structure/complex. To achieve this, we need  plasma parameters that satisfy eqns. \ref{eqn:dm_rm_limits}, \ref{eqn:lensing}  and \ref{eqn:tau_ff} simultaneously. A scan through the parameter space shows that self-consistent solutions are only obtained for $\eta_B<0.2$. If we further require  the putative filament to lie within the synchrotron nebula ($d_{\rm ls}<0.35\,$pc, as defined by the persistent source), then $\eta_B\lesssim 10^{-4}$ which will lead to unrealistically large magnetic fields. We therefor conclude that the same filamentary complex cannot provide the postulated plasma-lensing, and the observed Faraday rotation and variations thereof.

One could decouple Faraday-rotating and lensing plasma and readily find self-consistent solutions for two different plasma structures using eqns. \ref{eqn:dm_rm_limits} and \ref{eqn:lensing} separately. For instance, lensing can be caused by structures with plasma parameters of $T_4\sim 10^5$, ${\rm DM}_{10}\sim 1$, $n_e\sim 10^6\,{\rm cm}^{-3}$. The RM variations can be provided by transverse velocity of $\sim 30\,{\rm km/s}$ across a plasma structure with parameters of $T_4\sim 10^2$, $l\sim 2\times 10^{-4}\,{\rm pc}$ and $n_e\sim 5\times 10^4\,{\rm cm}^{-3}$.

\section{Conclusions}
\label{sec:conclusions}
We have shown that Faraday conversion of linear to circular polarised radiation is a relevant effect for Fast Radio Bursts that propagate through dense and relativistic magneto-ionic media. For our test case of FRB\,121102, we conclude the following.
\begin{enumerate}
    \item If the radio bursts pass through the synchrotron nebula then the latter must be an ad-mixture of highly relativistic and cold electrons. Specifically, if the electron energies are power law distributed then they must extend to $\gamma_{\rm min}\lesssim 3$ in order to not violate Faraday conversion limits.
    
    \item The `one-zone' family of magnetar models posited by \citet{margalit18} are only consistent with Faraday conversion constraints of the magnetic flux diffusion from the magnetar initiates almost immediately after its birth. For their benchmark model of $B=0.25\,$G, $N_\gamma\propto \gamma^{-1.3}$, $\gamma_{\rm inj} = 200$, age of 12.4\,yr, the onset must be within a month of magnetar's birth.
    
    \item In models where the persistent source associated with FRB\,121102 is confined by a colder Faraday-rotating plasma shell \citep[for e.g. ][]{waxman2017}, the latter is required to be shock-heated to $\sim 10^{7.5}-10^9\,$K, have an electron column of $10-17.5\,{\rm pc}\,{\rm cm}^{-3}$, a geometric parameter of $0.4\lesssim \eta_B\leq 1$, and magnetic field of $10\,{\rm mG}<\eta_B B < 18\,{\rm mG}$. The existence of such a self-consistent and over-constrained solution is not trivial and lends credence to the model.
    
    \item In models involving dense filaments as in the Crab nebula, the magneto-ionic variations (DM and RM) cannot come from the same plasma structures that also act as a plasma lens which is postulated  to generate certain time-frequency structures seen in FRB\,121102 \citep{cordes2017}.  
\end{enumerate} 
We emphasise that observations targeting the detection of Faraday-converted circular polarisation at $\sim 1$\,GHz in bursts from FRB\,121102 are likely of great interest. We further advocate for `de-rotation' circular polarised signals in FRBs with linear polarisation fractions below unity, in the event that Faraday-converted circular polarisation has been averaged out. Finally, we anticipate that the arguments presented here will be of significance to other FRBs (particularly of a repeating nature) that are associated with radio-synchrotron sources and/or dense magneto-ionic media.

\section*{Acknowledgements}
The authors thank (i) Andrei Grizinov and Eli Waxman for discussions on the implications of Faraday conversion, (ii) Eli Waxman for commenting on the manuscript, (iii) Jason Hessels for discussions regarding observational aspects of FRB\,121102, (iv) Dongzi Li for pointing out an error in our application of Farday conversion in an earlier version of the manuscript, and (v) the organisers of the 2018 Schwartz/Reisman Institute for Theoretical Physics workshop on Fast Radio Bursts for their hospitality.  Numerical computations used \texttt{scipy}, \texttt{numpy}, \texttt{Python2.7}, \texttt{Python3.0} and \texttt{matplotlib} was used to render figures.

\bibliographystyle{mnras}

\begin{thebibliography}{}
\bibitem[Backer et al.(2000)]{backer2000} Backer, D.~C., Wong, T., \& Valanju, J.\ 2000, \apj, 543, 740 
\bibitem[Bower et al.(2002)]{bower2002} Bower, G.~C., Falcke, H., Sault, R.~J., \& Backer, D.~C.\ 2002, \apj, 571, 843 
\bibitem[Chatterjee et al.(2017)]{chatterjee2017} Chatterjee, S., Law, C.~J., Wharton, R.~S., et al.\ 2017, \nat, 541, 58 
\bibitem[Cordes et al.(2017)]{cordes2017} Cordes, J.~M., Wasserman, I., Hessels, J.~W.~T., et al.\ 2017, \apj, 842, 35 
\bibitem[Gajjar et al.(2018)]{gajjar2018} Gajjar, V., Siemion, A.~P.~V., Price, D.~C., et al.\ 2018, \apj, 863, 2 
\bibitem[Hessels et al.(2018)]{hessels2018} Hessels, J.~W.~T., Spitler, L.~G., Seymour, A.~D., et al.\ 2018, arXiv:1811.10748 
\bibitem[Homan et al.(2001)]{hofman2001} Homan, D.~C., Attridge, J.~M., \& Wardle, J.~F.~C.\ 2001, \apj, 556, 113 
\bibitem[Huang \& Shcherbakov(2011)]{huang2011} Huang, L., \& Shcherbakov, R.~V.\ 2011, \mnras, 416, 2574 
\bibitem[Kennett \& Melrose(1998)]{kennett1998} Kennett, M., \& Melrose, D.\ 1998, \pasa, 15, 211 
\bibitem[Lorimer et al.(2007)]{lorimer2007} Lorimer, D.~R., Bailes, M., McLaughlin, M.~A., Narkevic, D.~J., \& Crawford, F.\ 2007, Science, 318, 777 
\bibitem[Marcote et al.(2017)]{marcote2017} Marcote, B., Paragi, Z., Hessels, J.~W.~T., et al.\ 2017, \apjl, 834, L8 
\bibitem[Margalit \& Metzger(2018)]{margalit18} Margalit, B., \& Metzger, B.~D.\ 2018, \apjl, 868, L4
\bibitem[Masui et al.(2015)]{masui2015} Masui, K., Lin, H.-H., Sievers, J., et al.\ 2015, \nat, 528, 523 
\bibitem[Michilli et al.(2018)]{michilli2018} Michilli, D., Seymour, A., Hessels, J.~W.~T., et al.\ 2018, \nat, 553, 182 
\bibitem[Pacholczyk \& Swihart(1970)]{pachol1970} Pacholczyk, A.~G., \& Swihart, T.~L.\ 1970, \apj, 161, 415 
\bibitem[Readhead(1994)]{readhead1994} Readhead, A.~C.~S.\ 1994, \apj, 426, 51 
\bibitem[Rybicki \& Lightman(1979)]{rybicki_lightman} Rybicki, G.~B., \& Lightman, A.~P.\ 1979, New York, Wiley-Interscience, 1979.~393 p., 
\bibitem[Sazonov(1969)]{sazonov1969} Sazonov, V.~N.\ 1969, \sovast, 13, 396 
\bibitem[Spitler et al.(2014)]{spitler2014} Spitler, L.~G., Cordes, J.~M., Hessels, J.~W.~T., et al.\ 2014, \apj, 790, 101 
\bibitem[Spitler et al.(2016)]{spitler2016} Spitler, L.~G., Scholz, P., Hessels, J.~W.~T., et al.\ 2016, \nat, 531, 202 
\bibitem[Tendulkar et al.(2017)]{tendulkar2017} Tendulkar, S.~P., Bassa, C.~G., Cordes, J.~M., et al.\ 2017, \apjl, 834, L7 
\bibitem[Thornton et al.(2013)]{thornton2013} Thornton, D., Stappers, B., Bailes, M., et al.\ 2013, Science, 341, 53 
\bibitem[Walker(1998)]{walker1998} Walker, M.~A.\ 1998, \mnras, 294, 307 
\bibitem[Waxman(2017)]{waxman2017} Waxman, E.\ 2017, \apj, 842, 34 

\end{thebibliography}

\label{lastpage}
\end{document}